\documentclass[traditabstract]{aa} 

\usepackage{graphicx}
\usepackage[numbers]{natbib}
\def\simgreat{\mathbin{\lower 3pt\hbox
     {$\rlap{\raise 5pt\hbox{$\char'076$}}\mathchar"7218$}}}
\def\simless{\mathbin{\lower 3pt\hbox
     {$\rlap{\raise 5pt\hbox{$\char'074$}}\mathchar"7218$}}}

\newcommand{\Lsun} {L$_{\odot}$}
\newcommand{\Msun} {M$_{\odot}$}

\begin{document}

\title{Shadows, gaps, and ring-like structures in protoplanetary
  disks}

\titlerunning{Shadows, gaps, and ring-like structures in protoplanetary disks}

\author {R.~Siebenmorgen\inst{1} \and F.Heymann\inst{1,2}}
\institute{
        European Southern Observatory, Karl-Schwarzschild-Str. 2,
        D-85748 Garching b. M\"unchen, Germany
\and
Department of Physics and Astronomy, University of Kentucky,
Lexington, KY 40506-0055, USA
}
\offprints{rsiebenm@eso.org}
\date{Received November 21, 2011 / Accepted December 30, 2011}

\abstract {We study the structure of passively heated disks around T
  Tauri and Herbig Ae stars, and present a vectorized Monte Carlo dust
  radiative transfer model of protoplanetary disks. The vectorization
  provides a speed up factor of $\sim$\,100 when compared to a scalar
  version of the code. Disks are composed of either fluffy carbon and
  silicate grains of various sizes or dust of the diffuse ISM.  The IR
  emission and the midplane temperature derived by the MC method
  differ from models where the radiative transfer is solved in slab
  geometry of small ring segments.  In the MC treatment, dusty halos
  above the disks are considered. Halos lead to an enhanced IR
  emission and warmer midplane temperature than do pure disks. Under
  the assumption of hydrostatic equilibrium we find that the disk in
  the inner rim puffs up, followed by a shadowed region.  The shadow
  reduces the temperature of the midplane and decreases the height of
  the extinction layer of the disk. It can be seen as a gap in the
  disk unless the surface is again exposed to direct stellar
  radiation. There the disk puffs up a second time, a third time and
  so forth. Therefore several gaps and ring--like structures are
  present in the disk surface and appear in emission images. They
  result from shadows in the disks and are present without the need to
  postulate the existence of any companion or planet.  As compared to
  Herbig Ae stars, such gaps and ring--like structures are more
  pronounced in regions of terrestrial planets around T Tauri stars.}


\maketitle

\section{Introduction}

Low--mass stars are surrounded by gaseous dust disks; prominent
examples are T Tauri and Herbig Ae/Be stars (Waters \& Waelkens,
1998).  The existence of disks is derived from observations of the
submillimeter continuum (Beckwith et al., 1990), the IR excess over
the photospheric emission (Kenyon \& Hartmann, 1995), and direct
imaging (McCaughrean \& O'dell, 1996).  Protoplanetary disks with weak
mid--IR and strong far--IR emission are called transitional disks
(Najita et al. 2007; Furlan et al. 2009; Cieza et al. 2008).  They are
of interest because the missing mid--IR emission could be caused by
grain growth or planet formation (Testi et al., 2003). Their optical
extinction is flatter than the standard ISM extinction curve
(Fitzpatrick \& Massa, 2007) and shows time variability (Schisano et
al. 2009).  This indicates that large grains or clumpy material orbits
and partially obscures the star.

The evolution of the disk is followed theoretically by studying the
time dependence of the surface density of gas and
dust. Hydro--dynamical models of viscous protoplanetary disks
including planets have been developed (Lubow et al. 1999; Ozernoy et
al. 2000; Klahr \& Bodenheimer 2003; Edgar \& Quillen 2008; Alexander
\& Armitage 2009). The models predict that the disk is cleared by an
orbiting planet and that the surface density shows structures with
holes and gaps, spiral arms, and rings. Such disk structures are
inferred from fitting the spectral energy distribution (SED) by
applying models considering pure disks (Calvet et al 2002; Espaillat
et al. 2010; Mulders et al. 2010) or disks with halos (Schegerer et
al. 2008; Verhoeff et al. 2011).

Recently, several disks with rings have been observed (van Boekel et
al. 2005; Fukagawa et al. 2006; Ratzka et al. 2007; Brown et al. 2007,
2008; Mayama et al. 2010; Thalmann et al. 2010; Olofsson et al. 2011).
In the gaps, between the rings, companion candidates are detected
(Huelamo et al. 2011; Kalas et al.  2008; Lagrange et
al. 2010). Hydro-models account for the observed radial distribution
of planets and lifetimes of the disks, which are a few Myr (Strom et
al. 1989; Bertout et al. 2007). The strength of the disk--planet
interactions are influenced by the gas pressure so that the structure
of protoplanetary disks plays a key role in understanding the planet
formation process.

There is a rich literature on radiative transfer models of
protoplanetary disks (e.g. Wolf et al. 2003; Whitney et al. 2003;
Pinte et al. 2006; Robitaille et al. 2006; Pinte et al. 2008). The
structure of the disks is discussed by Nomura (2002), Walker et
al. (2004), and Dullemond \& Dominik (2004a), and self--shadowing
ripples in the disks have already been mentioned by D'Alessio et
al. (1999) and Dullemond (2000); for a recent review, see Dullemond \&
Monnier (2010). In this paper the thermal structure of disks is
discussed by solving the hydrostatic and radiative balance
equations. We are interested in deriving the structure and IR
appearance of a starlight--heated, so--called passive disk. Dust
models used in this paper are described in Sect.~\ref{dust}, and a
vectorized three--dimensional Monte Carlo (MC) radiative transfer code
in Sect.~\ref{MC}. We compare results of the MC model against a method
that samples the disk in small ring segments at radius $r$ from the
star and in which the radiative transfer of the segments is solved in
a slab geometry (Sect.~\ref{1dvers3d}). The influence of different
dust opacities on the SED of the disks is discussed in
Sect.~\ref{sed.opac}.  We apply the models to T Tauri and Herbig Ae
stars. The effect of a dust halo on the emission and temperature
structure is presented (Sect.~\ref{halos}). The vertical temperature
structure introduces variations of the height of the disk surface
which causes shadows.  They emerge as gaps and ring--like structures
at distances from the star, which are thought to be the birthplaces of
terrestrial planets (Sect.~\ref{rings}). Strikingly the structures are
caused only by the coupling of the hydrostatic and radiative transfer
equations without the need to postulate a planet.

\section{Dust models \label{dust}}

In protoplanetary disks the properties of dust are thought to evolve,
and the mineralogy, composition, porosity and size distribution of the
grains differs from that of the diffuse ISM. Thanks to an enhanced
particle collision rate, fluffy grains are produced that grow in size
over time (Natta et al. 2004; Acke et al. 2004; Natta et al. 2007;
Lommen et al. 2007; Ricci et al. 2010). Disk particles settle towards
the midplane under gravity, a process counteracted by grain
fragmentation and disk turbulence (Mizuno et al. 1988; Weidenschilling
et al. 1993; Sterzik et al. 1994; Dullemond \& Dominik 2004b, 2005).
It is found that grains of radius $\simgreat 10\mu$m settle towards
the midplane, and depending on turbulent motions, will be blown up to
the surface, where they remain for long time periods, and well coupled
with the gas (Alexander \& Armitage 2007; Charnoz et al. 2011).

Dust gets annealed in high--temperature regions of the disk, and
crystalline structures are built.  They are detected and the observed
profiles of the 10$\mu$m silicate band differ from that of the
interstellar medium (Malfait et al. 1998; Bouwman et al. 2008;
Schegerer et al. 2006; Kessler--Silacci et al. 2006; Furlan et
al. 2011; Oliveira et al. 2010; Watson et al. 2009). About 4\% to
7.5\% of the dust mass of the Solar System is in crystalline silicates
whereas crystallization is $\sim 2$\% in the diffuse ISM (Kemper et
al., 2005).  Ice coatings are built in frosty regions (Siebenmorgen \&
Gredel 1997; Pontoppidan et al. 2005; Visser et al. 2009). The
location where water ice can exist is important for terrestrial planet
formation (Ida \& Lin 2008; Min et al. 2011). All of these processes
are combined by mixing and transport mechanisms and are valid for the
neutral part of the disks where grain charging can be ignored.
Ionization becomes important in the disk surface; therefore, the true
situation of the mineralogy in protoplanetary disks is a complicated
matter with a rich chemical network and evolution mechanisms in place
(Gail, 2004).

The SED and the thermal structure of the disks depend on the applied
dust model and for comparison we present results using dust opacities
as derived for protoplanetary disks and the diffuse ISM.  For the dust
in the ISM, we use a power--law size distribution: $n(a) \propto
a^{-3.5}$ with particle radii between $a_- \leq a \leq a_+$ of bare
spherical particles of silicates ($a_-=320\rm{\AA}$\,,
$a_+=2600\rm{\AA}$) with optical constants by Draine (2003) and carbon
($a_-=160\rm{\AA}$\,, $a_+=1300\rm{\AA})$, with optical constants by
Zubko et al. (1996), and both with bulk density of 2.5\,g/cm$^3$. We
use dust abundances (ppm) of 31 [Si]/[H] and 200 [C]/[H], which are in
reasonable agreement with cosmic abundance constraints (Asplund et
al., 2009). This gives a gas--to--dust mass ratio of 130. For
protoplanetary dust we consider fluffy agglomerates of silicate and
carbon as subparticles. Other parameters are as for the ISM dust
except that grains grow to 100 times larger radii of $a_+=33\mu$m. As
relative volume fraction of the composite grain we use 34\% silicates,
16\% carbon, and 50\% vacuum, which translates to a relative mass
fraction of 68\% silicates and 32\% aC. Absorption and scattering
cross-sections and the scattering asymmetry factor is computed with
Mie theory for the ISM grains, and we apply the Bruggeman rule for the
fluffy composites.  This gives a total mass extinction cross section
for the large grains in the optical (0.55$\mu$m) of $K^{ext}_V =
30600$ of the ISM dust and 4000\,[cm$^2$/g--dust] of the fluffy
composites. The wavelength dependence $K^{ext}_{\lambda}$ of both
models is similar to those displayed in Kr\"ugel \& Siebenmorgen
(1994, Fig.12).


\section{Vectorized MC radiative transfer \label{MC}}

The radiative transfer problem is solved by a MC technique by
following the flight path of many random photons. Basic ideas are
given by Witt (1977), Lucy (1999), and Bjorkman \& Wood (2001), and
the original version of our code was developed by Kr\"ugel (2008). Our
model space is a three--dimensional Cartesian grid $(x,y,z)$ that is
partitioned into cubes and, where a finer grid is required, further
divided into subcubes (cells).  The star of luminosity $L$ emits a
total of $N = n \cdot N_{\nu}$ photon packages per second and in each
of the $N_{\nu}$ frequency bins $n$ photon packages are emitted.  Each
package has a constant energy $\varepsilon = L/N$. A package entering
a cell on its flight path may be absorbed there or scattered.  The
probability of such an interaction is given by the extinction optical
depth along the path within the cell, $\Delta \tau$, and occurs when
$\Delta \tau \geq -log(\xi)$ using unified random number $\xi$.  The
package is scattered if the dust albedo $A > \xi'$, using random
number $\xi'$; otherwise, it is absorbed.  When the package is
scattered, it only changes direction determined in a probabilistic
manner by the phase function.  When it is absorbed, a new package of
the same energy, but usually different frequency $\nu'$ is emitted
from the spot of absorption.  The emission is isotropic.  Each
absorption event raises the energy of the cell by $\varepsilon$, and
accordingly its temperature.

The computational speed of the code is increased considerably by
calculating the flight paths of photons simultaneously.  This type of
vectorization is realized in two flavors using either conventional
computer processing units (CPU) within FORTRAN~90 and OMP
environment\footnote{www.openmp.org} or graphics processing units
(GPU) with NIVIDA cards and CUDA\footnote{www.nivida.com} (Heymann,
2010). The speed--up scales almost linearly with the number of
processing cores available.  For parallelization particular attention
needs to be paid to the random number generator, for which we choose
the Mersene Twister algorithm (Matsumoto \& Nishimura, 1998).

Verification of the vectorized MC code against benchmarks (Ivezic et
al. 1997; Pascucci et al. 2004) was done in an accompanying paper
(Heymann \& Siebenmorgen, 2012).  The code is also tested against the
spheric symmetrical ray tracing code by Kr\"ugel (2008). The SED is
computed by counting the photon packets leaving the cloud towards the
observer using a large beam. The observed intensity can also be
computed by following the radiative transfer of the line of sight from
the observer or a pixel of the detector plane, through the model
cloud using the MC computed dust temperatures and scattering events
(Heymann, 2010). The ray tracer is also vectorized and allows us to
derive dust scattering and emission images with high signal to noise, 
which is not possible by photon counting procedures.

The number of interactions of photon packets with the dust increases
exponentially with the optical depth of the cell. In cells of extreme
high optical depth, $\tau_{\rm V} \simgreat 1000$, the photon packages
are trapped and interact with the dust over and over again before they
have a chance to escape. In this way MC treatments are slowed down
considerably. Unfortunately, this situation appears in cells close to
the midplane of protoplanetary disks. A modified random walk (MRW)
procedure for improving the computational efficiency of MC methods has
been presented by Fleck \& Canfield (1984). The MRW enlarges the mean
free path length of packets by a diffusion approximation whenever
necessary, and has been tested by Min et al. (2009) and Robitaille
(2010). Let $r$ be the distance of the interaction point to the
nearest site wall of a cell. The MRW is considered in a cell when the
photon package has a distance $r \geq {\gamma}/\rho K_{R}$, where
$\gamma > 5$ is a user specified constant, $\rho$ the dust density,
and $K_{R}$ the Rosseland mean extinction coefficient. The photon
travel distance $d$ is derived from

\begin{equation}
  d = - \frac{1}{D} \left(\frac{r}{\pi}\right)^2   \rm{ln} \ \delta
\end{equation}

\noindent
with diffusion constant $D = 1/3 \rho K_{R}$ and a pre--computed
sample of $\delta$ given by

\begin{equation}
\xi = 2 \sum_{i=1}^{\infty} (-1)^{i+1} \left(\delta^{i}\right)^{2}
\end{equation}

\noindent
with unified random number $\xi$. The number of absorption events
within a cell is used to compute the dust temperature and, within $r$
of the MRW, is estimated by $n_{abs} = d\ \rho\ K_P$, where $K_{P}$ is
the Planck mean mass extinction coefficient of the dust. Ignorance of
these acceleration methods has a tremendous effect on the run time
requirements of a vectorized MC treatment. In our scheme the
trajectory of photon packages through the model space is vectorized in
so--called threads. Trajectories that hit the cells of high optical
depth have many more interactions than others, most likely the
majority of trajectories. This results in a rather unbalanced workload
over all threads, so that the advantage of vectorization is lost.  In
such cases the run time increases by the number of CPU or GPU cores
available, which in our case are 8 and 480, respectively. This factor
of efficiency loss is on top of the speed up gain of 5 -- 10 of the
MRW procedure achieved in scalar MC treatments (Pinte et al., 2009).

\begin{table*}[htb]
\caption{\label{para.tab} Parameters of the fiducial T Tauri and Herbig Ae disks.}
\begin{center}
\begin{tabular}{|l  l | l | l|}
\hline
Parameter &  & T Tauri  &Herbig Ae  \\ 
 & & & \\
\hline
Stellar luminosity       &$L_*$ [\Lsun]      & 2      & 50  \\ 
Stellar mass             &$M_*$ [\Msun]      & 1      & 2.5  \\ 
Photospheric temperature &$T_*$ [K]          & 4,000  & 10,000\\
Column density & 
$\Sigma(r) = \frac{\tau_{\perp}(\rm{1AU})}{K_{\rm{V}}} (\frac{r}{\rm{AU}})^{\gamma}$
\,[g--dust/cm$^2$] & \multicolumn{2}{c|}{$r < 1$\,AU: \quad $\gamma = 0.5$} \\
& &  \multicolumn{2}{c|}{$r \geq 1$\,AU: \quad $\gamma = -1$} \\
Vertical optical depth & $\tau_{\perp}(\rm{1AU})$  & \multicolumn{2}{c|}{10,000}\\
Dust density in halo & $\rho_{\rm{halo}}$\,[g-dust/cm$^3$] & \multicolumn{2}{c|}{$0$ or $1.5 \times 10^{-18}$} \\
Inner disk radius        &$r_{\rm{in}}$ & \multicolumn{2}{c|}{evaporation} \\
Outer disk radius        &$r_{\rm{out}}$ [AU] &  22.5 & 40 \\
\hline
\end{tabular}
\end{center}
\end{table*}

\section{Protoplanetary disk models \label{disks}}

The time scale for disks to attain thermal equilibrium is set by the
balance between heating and cooling. This time scale is much shorter
than the evolution of the disk or the heating source.  The number
density of the gas, even in the upper layers of the disk, exceeds
$10^5$\,cm$^{-3}$, so that gas and dust are thermally coupled. Matter
will spiral in the disk towards the star and dissipate gravitational
energy. For a thin disk that extends to the stellar surface, an
accretion luminosity of about a quarter of the stellar luminosity is
converted in this way into heat.  For classical T Tau stars, observed
accretion rates are $\sim 10^{-7...-9}$\Msun/yr (Muzerolle et al.,
1998). At early epochs when the accretion is strongest, dissipation
dominates, but later heating by stellar radiation becomes more
important. As fiducial model of the T Tau star, we take a mass of
1\Msun\/, a luminosity of 2\Lsun \/, and a photospheric temperature of
4000\,K.  For comparison we treat a disk heated by a Herbig Ae star
with 2.5\Msun \/, 50\Lsun \/ and 10$^4$\,K (van den Ancker et al.,
1997).

>From near--IR interferometry (Millan-Gabet et al., 2006) of T Tauri
and Herbig Ae stars, it has been found that the inner disk scales with
stellar luminosity $r_{\rm {in}} \propto \sqrt L$.  Such a dependency
is expected assuming that $r_{\rm {in}}$ is set by evaporation of
grains at temperatures of $1000 - 1500$\,K (Dullemond \& Monnier,
2010). For Herbig Be stars, larger deviations of the simple relation
are measured at $L > 10^3$\Lsun \/. Spatially resolved spectroscopy,
at milliarcsec resolution, allows detection of a hotter gaseous
emission component closer to the star (Eisner et al. 2009; Najita et
al. 2009; Pontoppidan et al., 2011).  Evaporation temperatures of
silicate and carbon particles are around 1500\,K. The grains in the
disk are assumed to be fluffy and to have been formed by coagulation
of much smaller compact interstellar grains.  The fluffy agglomerates
are held together by weak van der Waals forces (binding energies
$\simless 0.1$\,eV), and they are expected to fall apart into their
refractory constituents when the temperature of the composite grain
exceeds 1000\,K. The constituent particles will then be hotter than
the average porous grains because they are much smaller. We assume
that the disk extends inwards up to the point where porous grains
reach 1000\,K, or equivalently small interstellar grains would be at
about 1500\,K.

In star--forming regions dust is heated to about 30\,K. At a large
distance from the star grain heating by the ambient radiation field of
nearby low or massive stars becomes important. We do not make
additional assumptions on the outer radiation field and consider only
the primary central heating source. Models are computed to an outer
radius of $r_{\rm {out}} = 22.5$\,AU for T Tau and 40\,AU for Herbig
Ae disks where the midplane temperature drops below 30\,K.

Initially we assume that the disk is isothermal in $z$, and the density
is given by

\begin{equation}
 \rho(r,z) = \sqrt{\frac{2}{\pi}} \ \frac{\Sigma(r)}{H(r)} \ e^{- z^2/2 H^2}
\label{rhothermal}
\end{equation}

\noindent
with scale height $H^2 = k T_{\rm {mid}} r^3 /G M_*m$, surface density
$\Sigma(r)$, molecular mass $m = 2.3 m_p$ and midplane temperature,
$T_{\rm {mid}}$, for which we use a power law as initial guess.  The
height of the disk is set to $z_0 = 4.5 \ H$.  In models with pure
disks, the density $\rho_{\rm{halo}}$ above $z > z_o$ is 0 and
constant when a halo is considered (Table~\ref{para.tab}). The surface
density is adjusted to reach a given optical depth in the vertical
direction $\tau_{\perp} = \Sigma(r) K_{\rm V}$, with visual extinction
$K_{\rm V}$.  We use $\tau_{\perp} = \tau_{\rm {1AU}} \ (r/{\rm
  {AU}})^{\gamma}$ with $\gamma =-1$ for $ r \geq 1AU$ and $\gamma
=0.5$ otherwise (Min et al. 2011). For the porous grain model with
$a_+ = 33\mu$m, we take a vertical optical depth from the surface to
the midplane at 1\,AU of $\tau_{\perp} = 10,000$. This choice holds
computational time requirements of the MC code within reasonable
limits.  It translates to a surface density of $\Sigma(\rm{1AU}) =
5$\,g-dust/cm$^2$ or a gas surface density, which is half the value
estimated for the minimum mass of the early solar nebulae (Hayashi,
1981).  Higher surface densities can be obtained considering larger or
ice--coated grains because $K_{\rm V}$ is reduced in such models
(Kr\"ugel \& Siebenmorgen, 1994).

\subsection{Disks with slab geometry}

A solution of the radiative transfer of a hydrostatic and
geometrically thin disk is developed by Kr\"ugel (2008, Sect.~11.3).
The disk is symmetric with respect to the midplane at $z=0$. The
density structure is given in cylindrical coordinates
$\rho(r,z)$. Light from the star falls on the disk under a grazing
angle $\alpha_{\rm{gr}}$ so that the star is visible from everywhere
on the disk surface (Armitage, 2007).  Flat disks with a constant
grazing angle of $\alpha_{\rm{gr}} = 2^{{\rm}{o}}$ and flared disks,
similar to Chiang \& Goldreich (1997), are considered with
$3^{{\rm}{o}} \leq \alpha_{\rm{gr}} \propto r^{2/7} \leq
7^{{\rm}{o}}$. The flat disk has a smaller grazing angle and
intercepts less stellar radiation than the flared disk, and even more
so at large distances from the star.

The disk is divided into small ring segments of width $\Delta r$ at
radius $r$ from the star.  For each ring the radiative transfer is
solved under all inclination angles $\theta$, for incoming $I^-$ and
outgoing $I^+$ radiation, assuming a slab geometry.  In the vertical
direction the opacities are so high, e.g.  $\tau_{\perp}{\rm {(1AU)}}
= 10^4$, that each ring segment is split into a completely opaque
midlayer and a much thinner top layer of optical thickness $\tau_{\rm
  {top}}$. The midlayer is assumed to be isothermal at temperature
$T_{\rm{mid}}$ and the top layer extends so far down that the
temperature at its bottom approaches $T_{\rm{mid}}$. For computing the
spectral energy distribution it is sufficient to choose $\tau_{\rm
  {top}} \sim 20$. The ray tracer is an iteration procedure and yields
the temperature structure $T(r,z)$ and the intensities of the upward
$I^+$ and downward $I^-$ directed radiation. By observing the disk at
viewing angle $\theta_{\rm{obs}}$, the received flux is computed from
the emission of the star, the intensity $I^+$ at the disk surface at
$\theta = \theta_{\rm{obs}}$ as well as by summing up contributions
from all rings. The radiative transfer is solved for each ring segment
at radius $r$ separately. However, the propagation of the radiation in
radial direction from one ring into the next is ignored. The procedure
is often called the {\it 1+1D} disk model.

\subsection{Disks with axial symmetry}

In disks that are in hydrostatic equilibrium in a vertical direction, the
gravitational force is balanced by the pressure gradient\footnote{The
  protoplanetary disk masses are well below the stellar mass
  $M_{\rm{disk}} \leq 0.01 M_{*}$.}:

\begin{equation}
- \frac{z}{r} \frac{GM_*}{r^2} = \frac{1}{\rho}\frac{dP}{dz}
\label{hydro}
\end{equation}

\noindent 
with pressure $P = \rho k T(z) / m$. For each height $z$ and radius
$r^2 = x^2+y^2$ from the star, the temperature $T(r,z)$ is computed
using the MC code of Sect.~\ref{MC}.  The star is set at
origin. Azimuthal dependencies in the disk structure are not
considered, so that the problem can be solved in axial symmetry.  It
allows us to solve the radiative transfer in one octant of a cube and
reduces the computational time by a factor 8 when compared to a full
three dimensional treatment.  For each height $z \geq 0$ and radial
distance at $x \geq 0$ and $y \geq 0$, we compute the azimuthal
average of absorbed photon packets. This average is used to derive the
dust temperature $T(r,z)$, and with Eq.~\ref{hydro} a new estimate of
the density structure of the disk $\rho(r,z)$. Obviously there is an
iterative scheme: starting with an initial guess of the midplane
temperature and a first set up of the disk structure $\rho(r,z)$ using
Eq.~\ref{rhothermal}, we compute the temperatures $T(r,z)$ after a MC
run. Once the temperatures are inserted into Eq.~\ref{hydro}, we
arrive at a new density structure $\rho(r,z)$, which is used to update
$T(r,z)$ with a new MC run. Typically after less than a dozen
iterations, the program converges to a stable disk configuration. The
problem we are solving is intrinsically two dimensional, but our MC
code is based on cubic cells that are three dimensional, therefore we
label such computations {\it {2D}} disk models hereafter.

Particular attention has to be paid to a good set--up of the grid used
in the MC code.  In the positive $(x,y,z)$ directions, we use a basic
grid of (350,350,50) cubes.  At locations where the radiation field
varies strongly, cubes are sampled into subcubes. Large gradients of
the radiation field are expected close to the surface, within the
extinction layer or at the inner wall of the disk. Estimates of the
height of the extinction layer $z_0$ and its thickness $\ell \sim H/2$
is given by Siebenmorgen \& Kr\"ugel (2010).

\begin{figure}[htp]
\begin{center}
\includegraphics[angle=0,width=8cm]{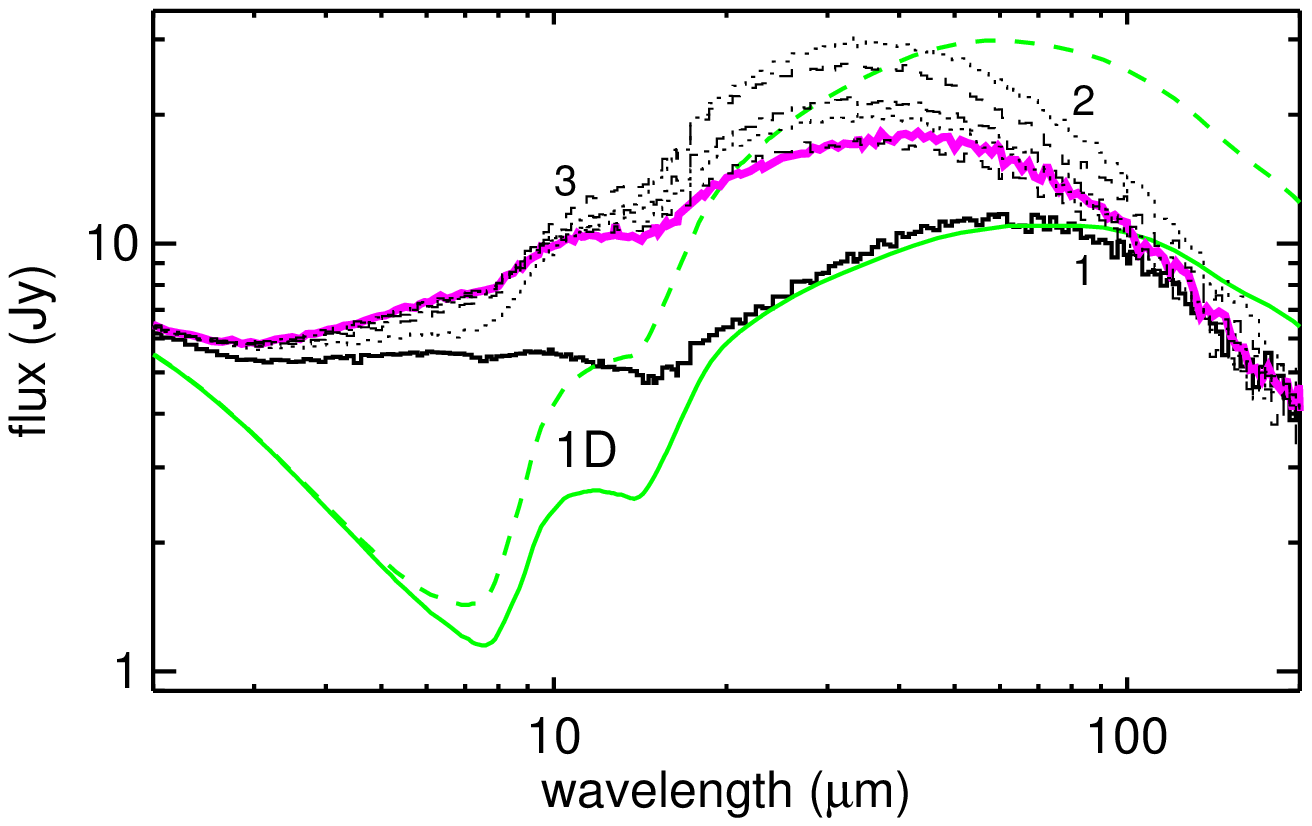}
\includegraphics[angle=0,width=8cm]{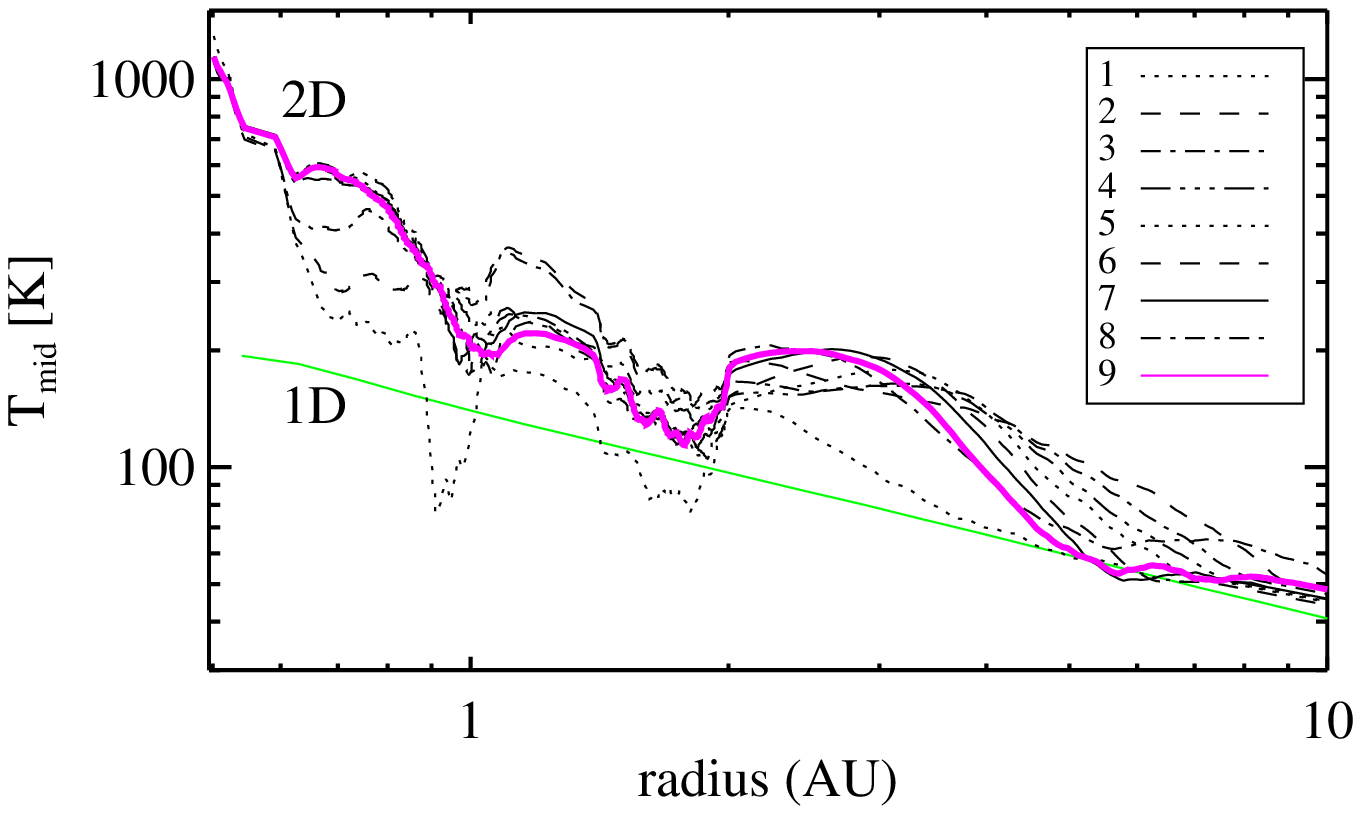}
\end{center}
\caption{{\it {Top}}: IR emission of a T Tau disk viewed at $\sim
  45^{\circ}$, which is either computed using the 1+1D model (label
  '1D') in a flat (green full line), a flaring (dashed green line)
  disk configuration, or the 2D disk.  Iterations 1 to 9
  of Eq.~(\ref{hydro}) of the 2D disk are shown. {\it
    {Bottom}}: Midplane temperatures of the 1+1D flat disk and the
  various iterations of the 2D disk. \label{sedIt.ps}}
\end{figure}

The extinction layer is sampled with cubes having $\tau_{\rm V}
\simless 0.6$. This is achieved by sampling cubes located close to the
extinction layer in up to 20 subcubes. In total the disk has $10^7$
cells, which provides a sampling with high spatial resolution. In each
MC run a number of $10^8$ photon packets are emitted from the
star. The fairly high number of packets provide a high--energy
resolution and is required to arrive at a stable estimate of the
temperatures $T(r,z)$, in particular near the midplane.

\subsection{Comparison: 1+1D versus 2D \label{1dvers3d}}

The SED and the midplane temperatures of a flat and flaring 1+1D disk
are compared with those of a 2D disk model. As an example, we take a T
Tauri star with parameters as specified in Table~\ref{para.tab}. For
ease of comparison, the distribution of the dust column density
$\Sigma(r)$ is chosen to be identical in all three models.  We take
the one computed by the 2D disk as evaporation radius.  The SED is
compared in the upper panel of Fig.~\ref{sedIt.ps}. In a 1+1D disk the
heating is proportional to the grazing angle. Since flaring disks have
a larger grazing angle than flat disks they show higher dust
re-emission luminosities and appear warmer. The assumption of 1+1D
disk models is that stellar photons impinge upon the disk at a given
radius with constant grazing angle. Such a simple picture fails close
to the dust evaporation zone. In early 1+1D models the inner disk
region was treated as an optically thick vertical wall at constant
temperature (Natta et al. 2001; Dullemond et al. 2001). Then the
dependency of gas pressure on the dust evaporation temperature is
included, resulting in a curved evaporation zone (Isella \& Natta
2005; Kama et al. 2009).  In the MC scheme the radiation transport in
radial direction is included, which, because of the fairly difficult
geometry, is otherwise not easy to implement in ray--tracing
techniques as used in 1+1D disks.  The SED of a 2D disk is very
distinct with a much stronger mid--IR emission and overall flatter
appearance. We also show in Fig.~\ref{sedIt.ps} the convergence of the
SED and the midplane temperature of the MC model as computed during
the first nine iterations of the density structure following
Eq.\ref{hydro}. The midplane temperature distributions of the 1+1D
disks are described well by a power law, whereas in the 2D disk
striking up and downs with radius are derived. We extensively tested
that the structure is not due to systematic effects that could be
introduced by the noise caused by a pure photon statistics or
inadequate sampling of the MC cells.  Following Eq.~\ref{rhothermal}
the midplane temperature is directly linked to the density, and one
expects a hilly path, as derived in the lower panel of
Fig.~\ref{sedIt.ps}, to be reflected in the overall observed disk
structure (Sect.~\ref{rings}).


\begin{figure}[htp]
\hspace{-1cm}
\includegraphics[angle=0,width=10cm]{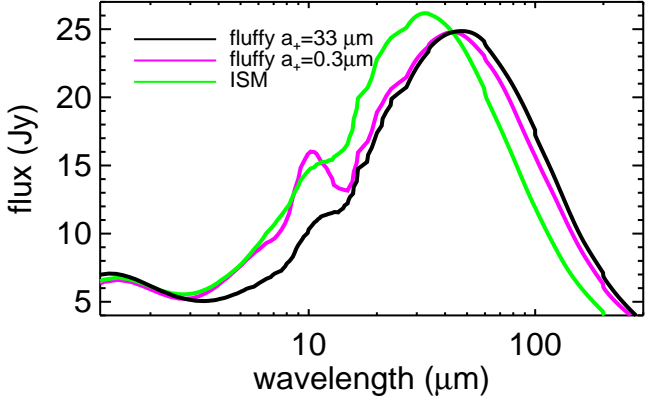}
\caption{IR emission of 2D disks viewed nearly face--on ($\sim
  20^{\circ}$) computed by applying different dust
  opacities. \label{sdust.ps}}
\end{figure}

\subsection{SED dependency on dust opacities \label{sed.opac}}

A realistic description of dust in protoplanetary disks is an open and
controversial issue, as are the applied dust absorption and scattering
efficiencies. Because these parameters directly enter into the
radiative transfer equation and the MC method, we wish to study their
impact on the SED.  As an example we take a T Tauri star with the
parameters in Table~\ref{para.tab}.  Three different dust models are
distinguished: a) ISM like grains, b) fluffy grains with particle
radii up to $a_+ = 0.3\mu$m, and c) fluffy grains with $a_+ =
33\mu$m. Other parameters remain as given in Sect.~\ref{dust}.  The SEDs
of the disks are compared using identical distributions of the column
density $\Sigma(r)$. Since the dust models have different visual
extinction coefficients $K_{\rm V}$, the vertical optical depth at 1AU
of $10^4$ for our standard dust (model c) needs to be increased by a
factor 7.7 for ISM dust and by a factor 8.5 for fluffy grains with
size distribution, as for ISM grains. SEDs of the 2D disks are
displayed in Fig.~\ref{sdust.ps}, where one notices that: 

i) The flux in the far--IR peaks at longer wavelength when applying
fluffy grains.

ii) The silicate emission profile depends on grain porosity, particles
size, and dust temperature (Voshchinikov \& Henning, 2008).  In the
band the cross section peaks at 9.5$\mu$m for ISM dust and at
9.8$\mu$m for the fluffy composites.  The feature--to--continuum ratio
of the extinction cross section in the 10$\mu$m band, when averaged
over the dust size distribution, is 25\% higher for ISM than for
fluffy grains of same size. Nevertheless, because of the detailed
temperature structure, model b), which is made of fluffy grains with
sizes typical of the ISM, shows the most striking silicate emission
feature. 

iii) The disk with fluffy grains up to $a_+ = 0.3\mu$m have a similar
submillimeter slope as the one with 100 times larger, $a_+ = 33\mu$m,
particles (Fig.~\ref{sdust.ps}).  This agrees with observations of
protoplanetary disks where the silicate feature is not correlated with
the millimeter slope of the SEDs (Lommen et al., 2010).

\begin{figure}[htp]
\includegraphics[angle=0,width=8.5cm]{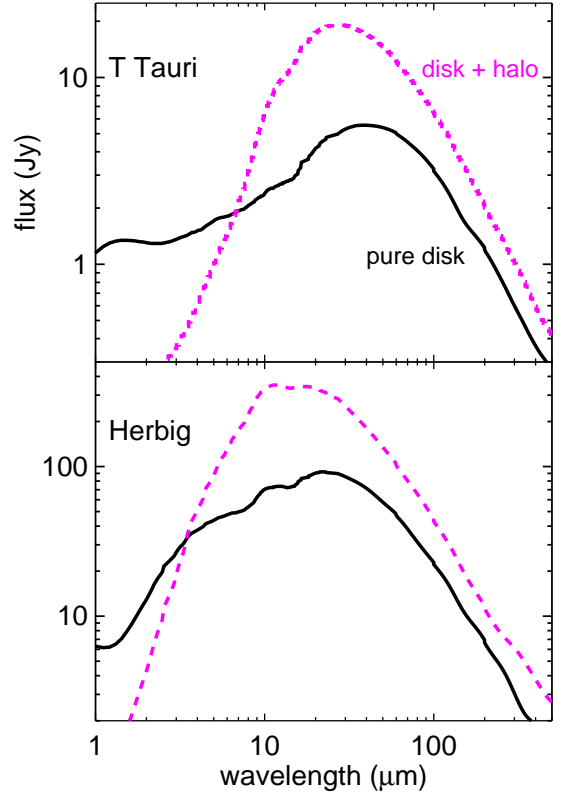}
\caption{IR emission of 2D disks viewed nearly edge--on ($\sim
  20^{\circ}$) with (dashed) and without (full line) a dusty halo.
  The disks are either heated by a T Tauri star (top) or a Herbig Ae
  star (bottom). \label{halo.ps}}
\end{figure}


\subsection {Disks plus halos \label{halos}}

Protoplanetary disks are thought to be formed from their initial
spheroidal configuration of the protostellar envelopes (Watson et al.,
2007). It is speculated that there is some residual dust left or that
it is replenished at high altitudes above the disk (Vinkovic et al.,
2006).  Dust in a halo decays through radiation pressure and gravity
and may be replenished by dynamical processes, such as outflows and
winds from the disk surface, or by planets in highly eccentric orbits
(Krijt \& Dominik, 2011).  Miroshnichenko et al. (1999) added the
emission of an optical thin halo onto the SEDs computed for optically
thick disks. They found that the halo dominates the IR emission and
provides additional disk heating on top of the direct stellar
radiation. The MC scheme allows a self--consistent treatment of the
radiative transfer in a configuration of a disk plus halo.  In the T
Tauri disk models (Table~\ref{para.tab}) we add a halo so that the
minimum dust density of Eqs.~\ref{rhothermal} and \ref{hydro} in each
cell of the MC grid exceeds $\rho(r,z) \geq \rho_{\rm{halo}} =1.5
\times 10^{-18}$\,(g-dust/cm$^3$).  This gives an optical depth for
the halo of $\tau_{\rm V} \sim 0.4$. The resulting SED of a disk plus
halo is shown in Fig.~\ref{halo.ps}. The SED appears warmer than the
pure disk with $\rho_{\rm{halo}}=0$. In the example, the disk plus
halo model produces a factor $\sim 3$ stronger IR peak emission and
less near--IR flux for wavelengths below $6\mu$m.  The halo provides
additional heating on top of the direct stellar light. Their midplane
temperature becomes a factor $\sim 2$ warmer than derived in a disk
without halo within 1 -- 10\,AU of the star.

We repeat the exercise for the more luminous Herbig Ae stars with the
parameters given in Table~\ref{para.tab}. With the same minimum dust
density, $\rho_{\rm{halo}}$, the optical depth of the halo is
$\tau_{\rm V} \sim 0.6$ as for the T Tauri star.  The resulting SEDs
are shown in Fig.~\ref{halo.ps}. The halo dominates the IR emission
for wavelengths $\simgreat 3\mu$m and produces warmer dust in the
midplane than in pure disks. The midplane temperature of the
Herbig Ae stars are displayed in Fig.~{\ref{HEBEsurf.ps}}.  The
midplane temperature in the pure disk shows a rapid decline in the
inner 3--4\,AU, followed by a $1/r^{0.6}$ dependency up to 12\,AU and
a strong decline farther out. The disk plus halo model is smoother and
fit by $T_{\rm {mid}} \sim 1/r$ in the 2--15\,AU region
(Fig.~{\ref{HEBEsurf.ps}}).


\begin{figure}[htp]
\begin{center}
\includegraphics[angle=0,width=9.4cm]{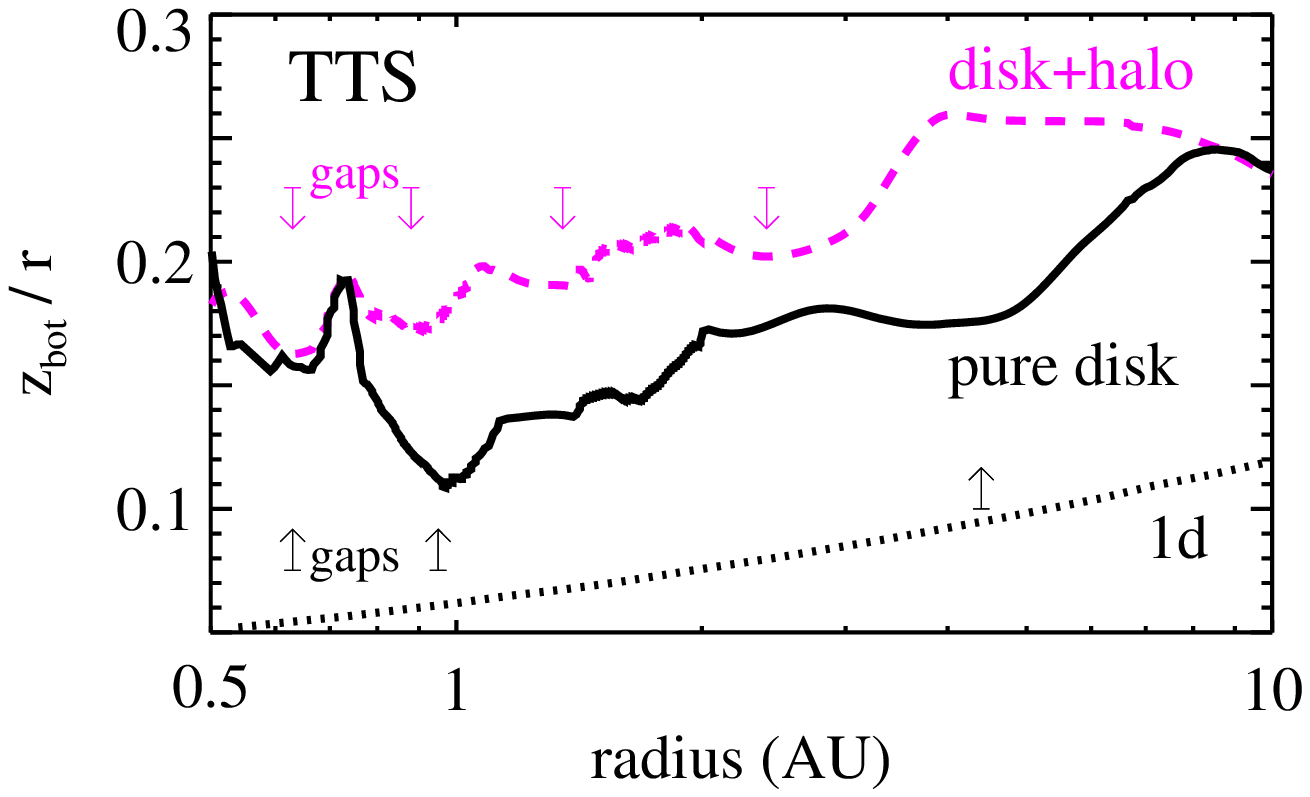}
\vspace{-0.1cm}
\includegraphics[angle=0,width=8cm]{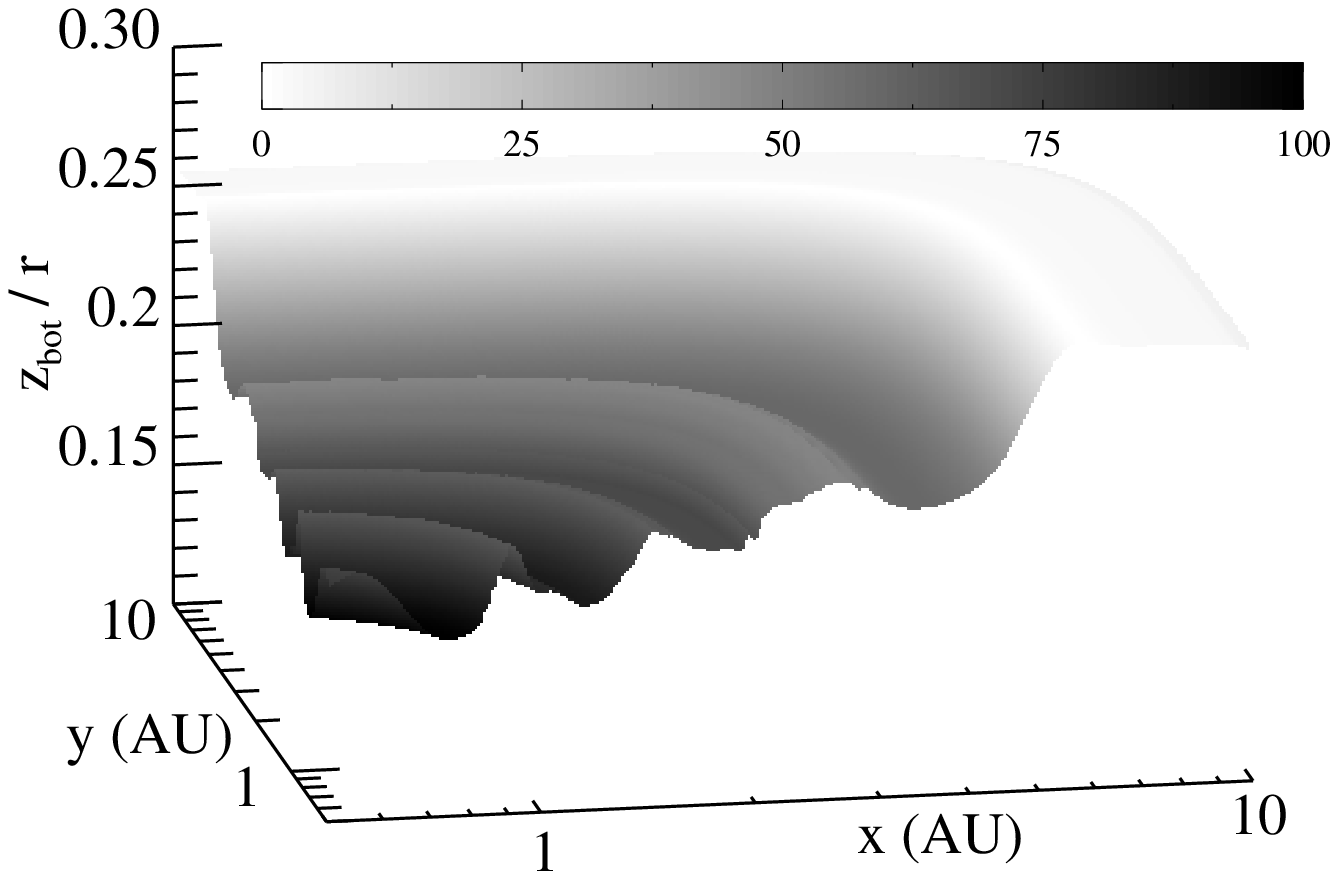}
\vspace{-0.1cm}
\includegraphics[angle=0,width=8cm]{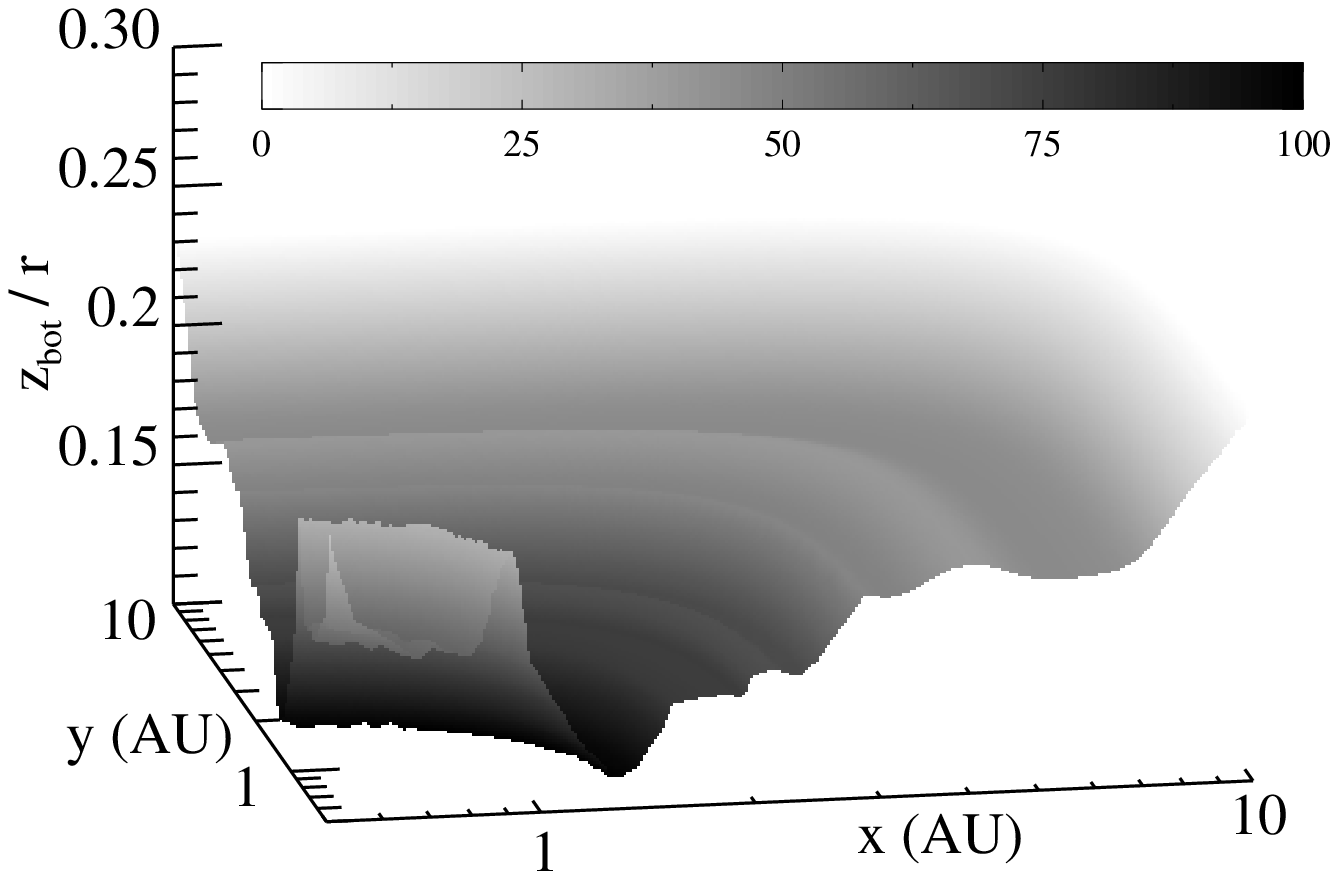}
\end{center}
\caption{ {\it {Top}}: The relative height of the disk photosphere,
  $z_{\rm {bot}}/r$, as a function of radius of a disk heated by a T
  Tauri star without (full line) and with (dashed) halo.  The scale
  height over radius, $H/r$, of a flat 1+1D disk (dashed) is shown for
  comparison. Gaps between ring structures are indicated. A shaded
  representation of $z_{\rm {bot}}/r$ along the (x,y)--plane is
  displayed for a disk plus halo configuration (middle) and a pure
  disk (bottom). Color bars give the relative gray scale (\%) in units
  of $z_{\rm {bot}}/r$. \label{TTSsurf.ps}}
\end{figure}

\section {Ring structure of protoplanetary disks \label{rings}}

We notice that the midplane temperature in 2D disks of T Tauri stars
show a wavy structure in particular near the zone where terrestrial
planets are supposed to form (Fig.~\ref{halo.ps}). Here we investigate
how this temperature behavior manifests itself in the appearance of
the disk.  First we study the height of the photosphere which we
define as the height of the bottom of the extinction layer,
$z_{\rm{bot}}$, where the vertical optical depth $\tau_{\perp}({\rm
  V}) = 1$.  In Fig.\ref{TTSsurf.ps} we show the height of the disk
photosphere as a function of radius, expressed as a unitless ratio
$z_{\rm{bot}}/r$.  We do this for the T Tauri disk models with and
without a halo using parameters of Table~\ref{para.tab}. For 1+1D
disks the midplane temperature is a smooth function declining with
radius, which can be approximated by a power law, and so is the scale
height which smoothly increases with distance (Fig.\ref{TTSsurf.ps}).
The geometrical thickness of the 1+1D disk increases with radius.

\begin{figure}[htp]
\begin{center}
\includegraphics[angle=0,width=6.5cm]{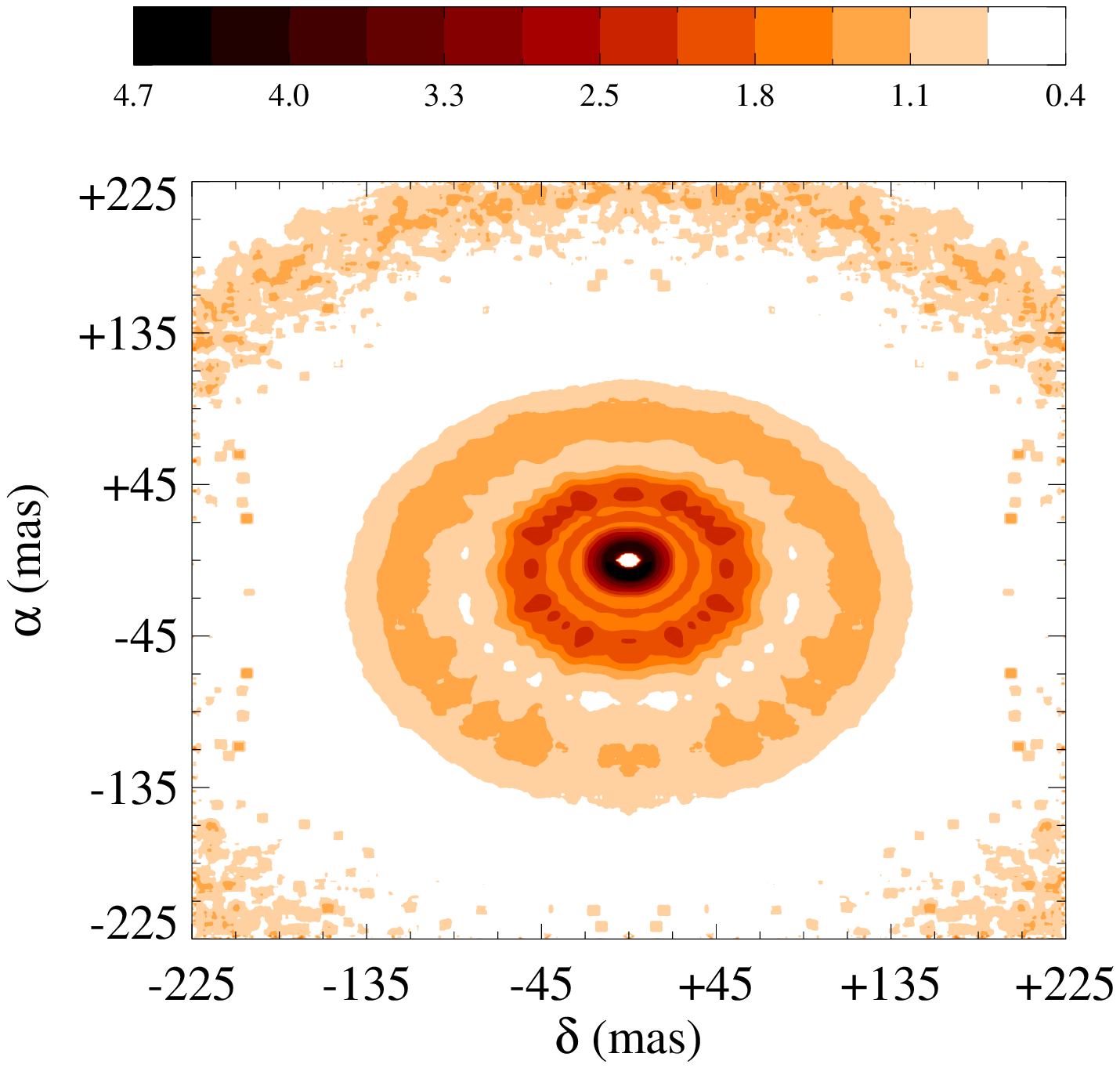}
\vspace{-0.6cm}
\includegraphics[angle=0,width=6.5cm]{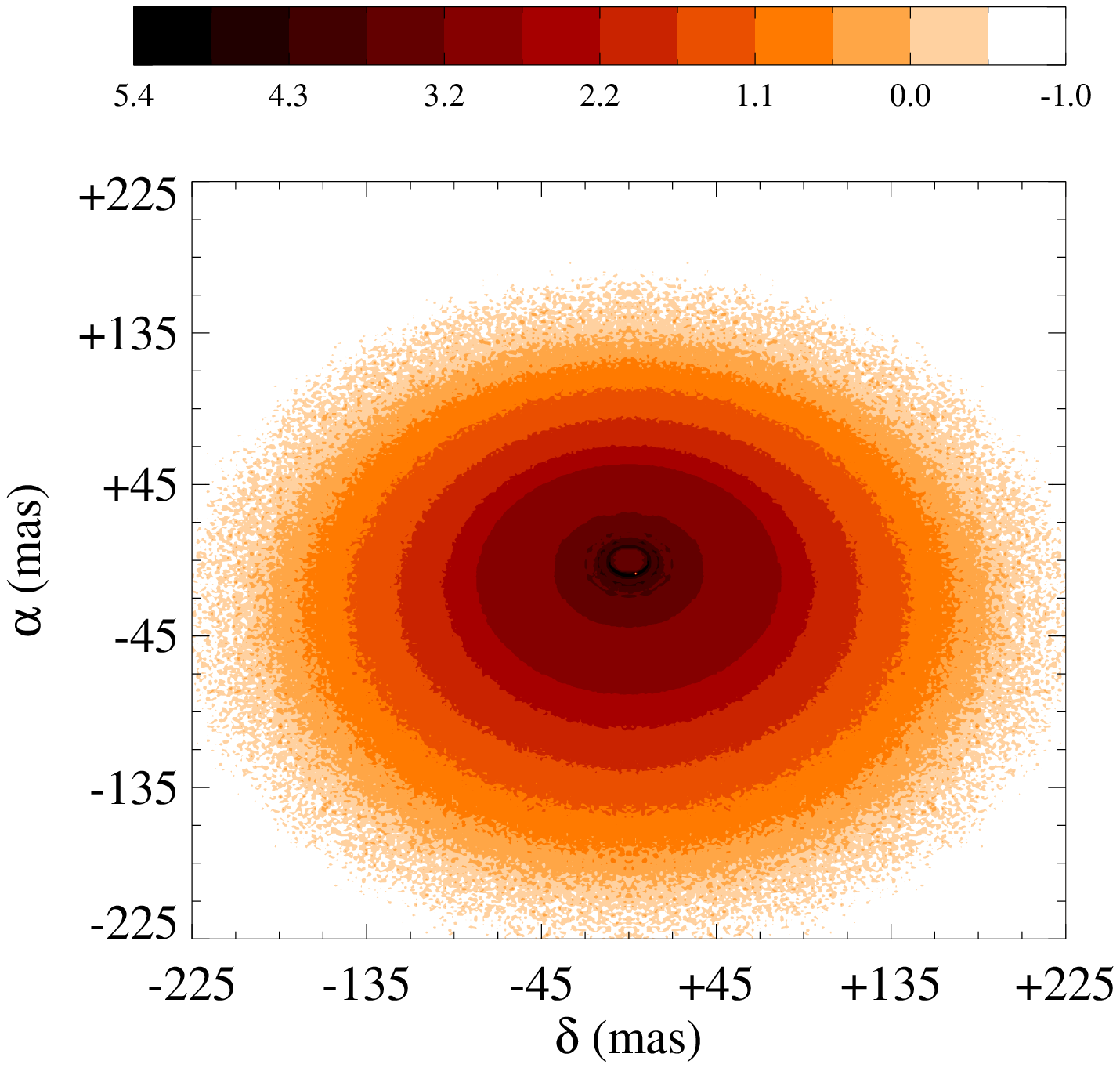}
\end{center}
\vspace{0.1cm}
\hspace{0.5cm}
\includegraphics[angle=0,width=6.5cm,height=5.cm]{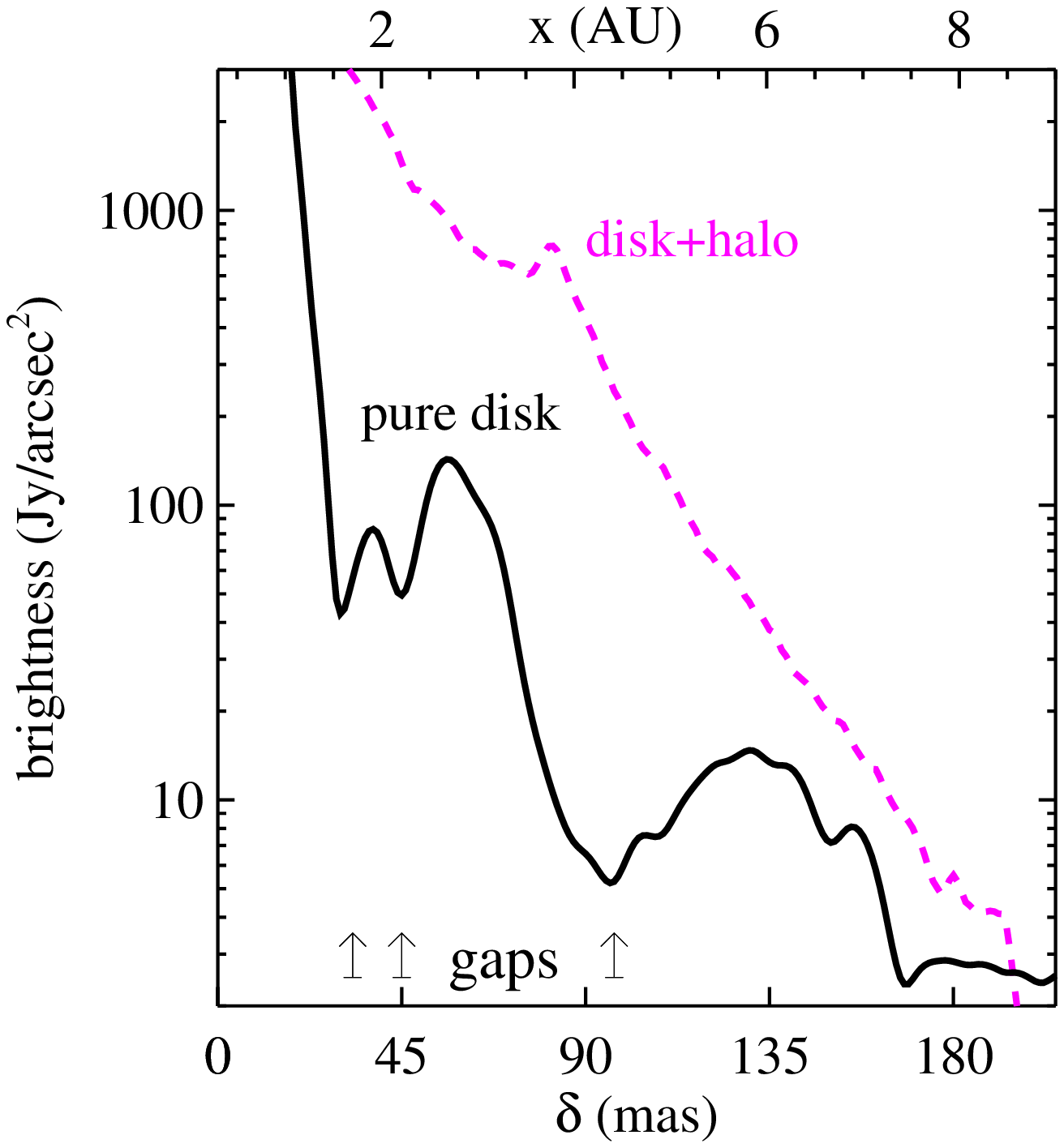}
\caption{Mid--IR images of disks heated by a T Tauri star at distance
  of 50\,pc and viewed at $30^{\circ}$. The image of a pure disk
  ({\it{top}}) and a disk plus halo is shown ({\it {middle}}). Color
  bars are in $\log($Jy/arcsec$^2)$.  {\it {Bottom}}: Corresponding
  surface brightness distribution measured as cut along $\delta$
  through origin of the image. Gaps between emission rings are
  indicated. \label{TTSima.ps}}
\end{figure}

\begin{figure}[htp]
\includegraphics[angle=0,width=7.7cm]{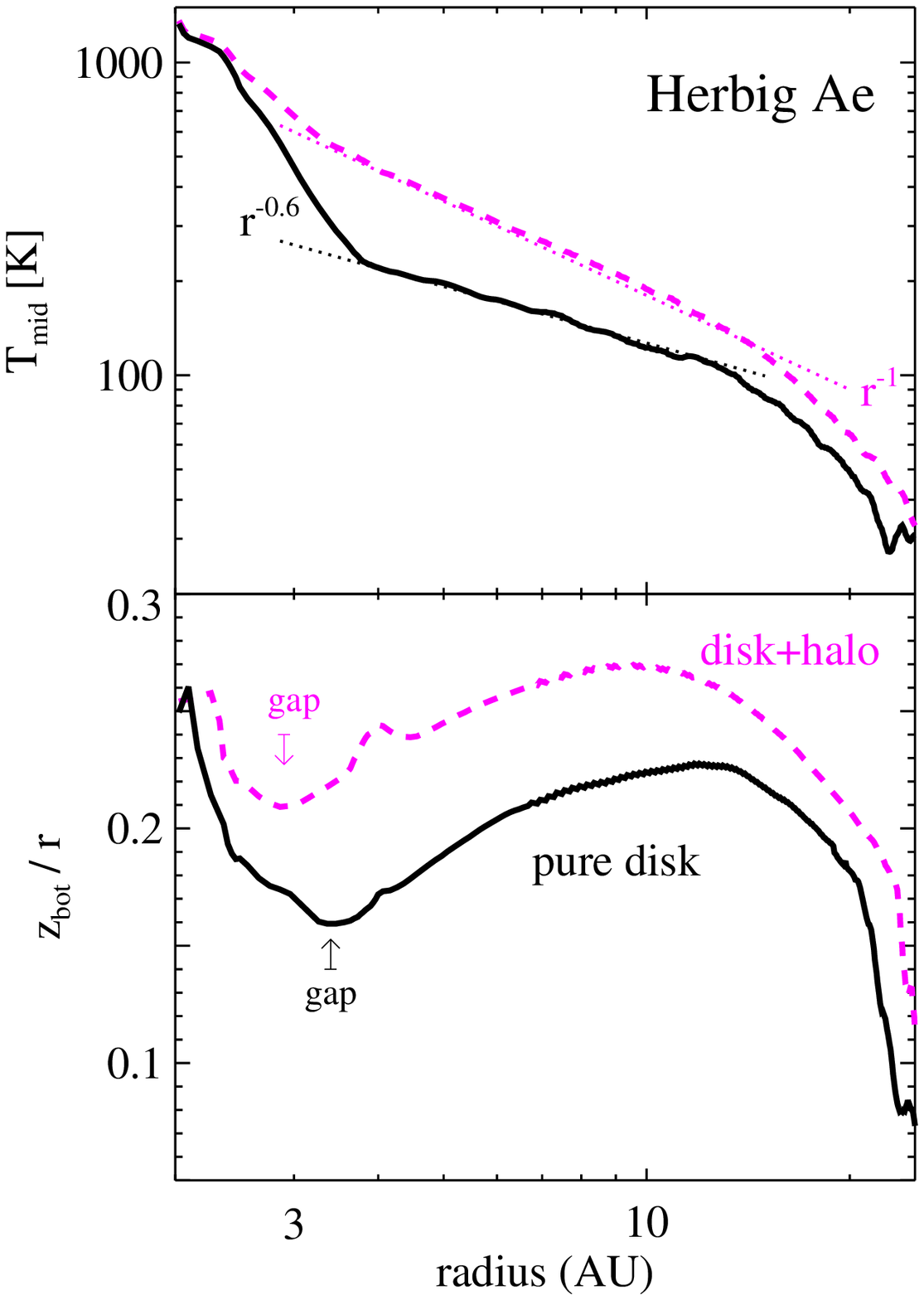}
\vspace{-0.05cm}
\includegraphics[angle=0,width=7.cm]{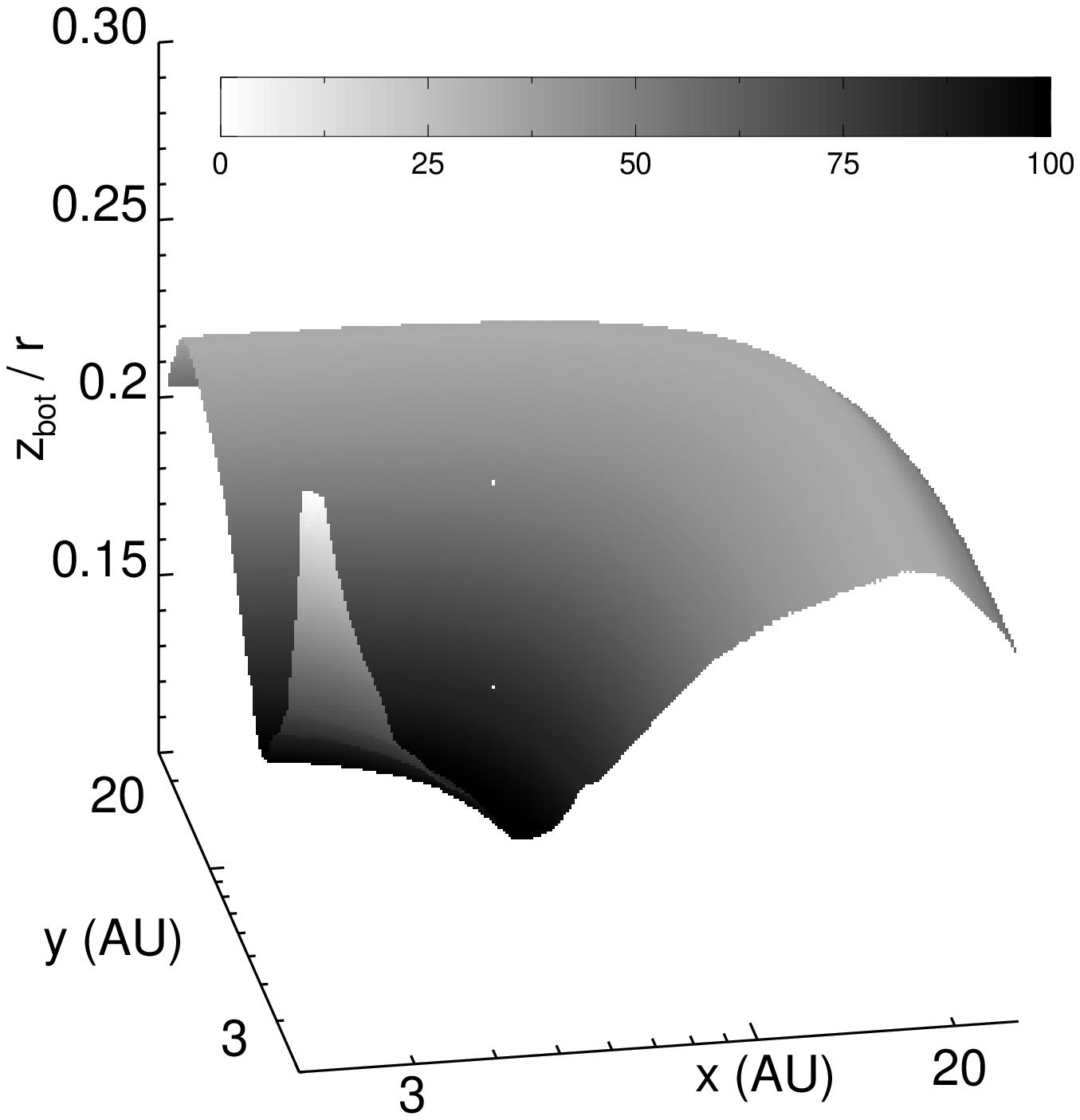}
\caption{{\it {Top}}: The midplane temperature and relative height of
  the disk photosphere, $z_{\rm {bot}}/r$, as a function of radius of
  a 2D disk heated by a Herbig Ae star.  Models are shown in a
  configuration of a pure disk (full line) and a disk plus halo
  (dashed). Power law fits (dotted) to the midplane temperature
  distribution are as labeled.  Gaps between ring structures are
  indicated.  {\it {Bottom}}: A shaded representation of $z_{\rm
    {bot}}/r$ along the (x,y)--plane is shown for a pure disk model.
  Color bar gives relative gray scale (\%) in units of $z_{\rm
    bot}/r$.
\label{HEBEsurf.ps}}
\end{figure}

The 2D disk near the evaporation zone is puffed up, and then its
thickness declines similar to the decrease in the midplane
temperature.  The temperature reduction is to be understand by
shadowed region of the disk surface where light from the star is
extincted and where grain heating becomes less efficient.  The shadow
is located close behind from the puffed--up inner rim. The disk becomes
thicker with increasing distance, because the gravity decreases, so
that there is a point where the shadow becomes less efficient, and the
disk is exposed to direct stellar light.  There again the disk is
puffed up followed by a second shadow, which is located at about 1\,AU
in our example and a third one near 4\,AU. Farther out ($\simgreat
10$\, AU) the midplane temperature drops to below 30\,K because the
stellar heating of the passive disk is inefficient, and another
distortion in the disk surface is not visible in most cases.  The
detailed structure of passively heated T Tauri disks are shown in
Fig.\ref{TTSsurf.ps} within the region of terrestrial planets. If one
considers halos, then gaps and ring--like structures are dimmed and
the disk appears thicker because their midplane temperature is
warmer. The disk surface is visualized in a three dimensional
representation by rotating the function $z_{\rm{bot}}(r)/r$ along the
midplane (Fig.~\ref{TTSsurf.ps}). The innermost puffed--up rim,
several gaps, and an overall wavy surface structure of the disk are
visible.  In Fig.~\ref{TTSima.ps} we display images and surface
profiles for the mid--IR. At other wavelengths the emission structure
is similar. In the image of the disk without halo there is a wide gap
opening farther out and smaller gaps visible closer to the star. This
structure is dimmed when a halo is considered.

\clearpage

\begin{figure}[htp]
\begin{center}
\includegraphics[angle=0,width=6.6cm]{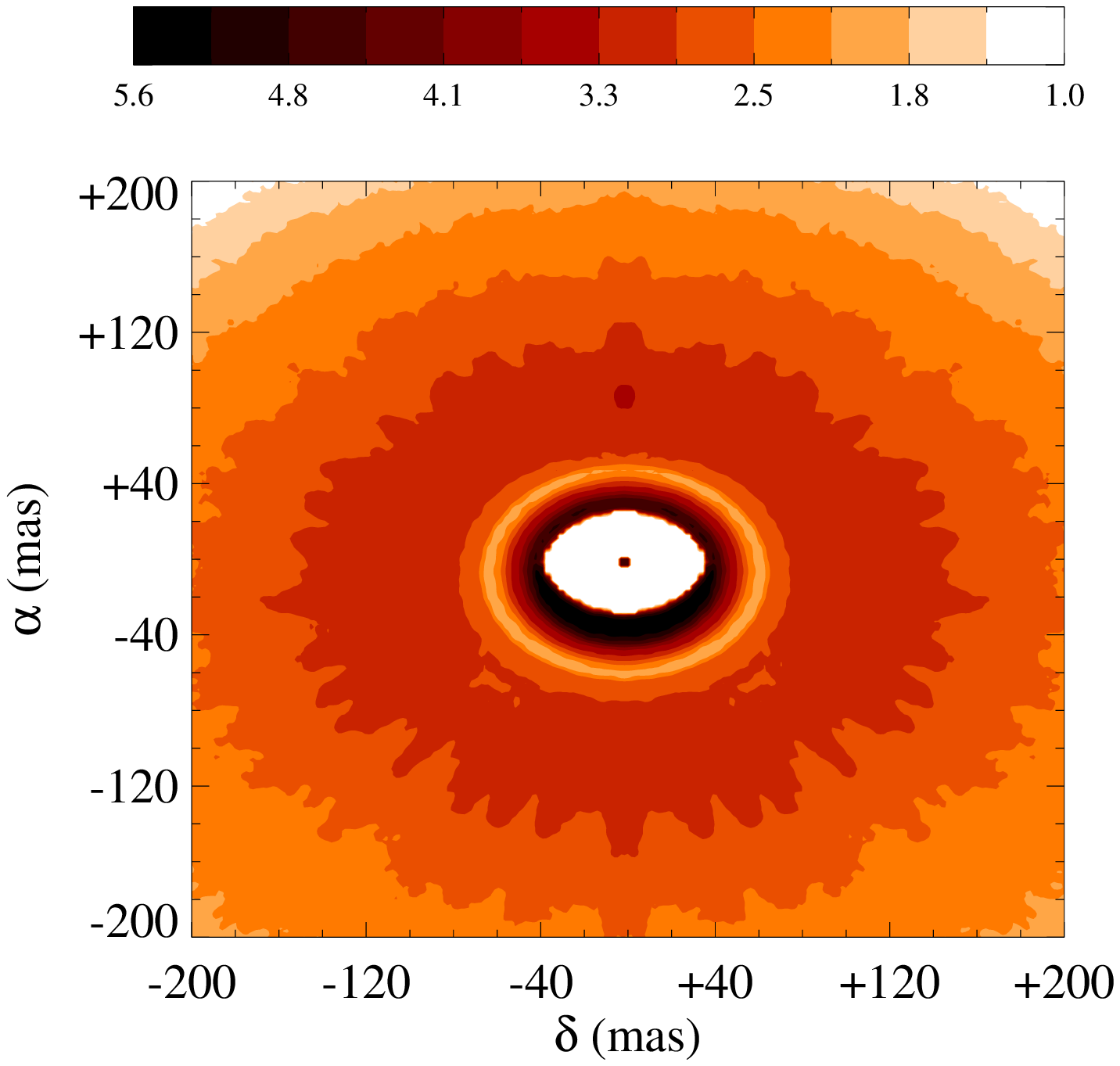}
\end{center}
\begin{center}
\hspace{-1.0cm}
\includegraphics[angle=0,width=6.8cm,height=5.cm]{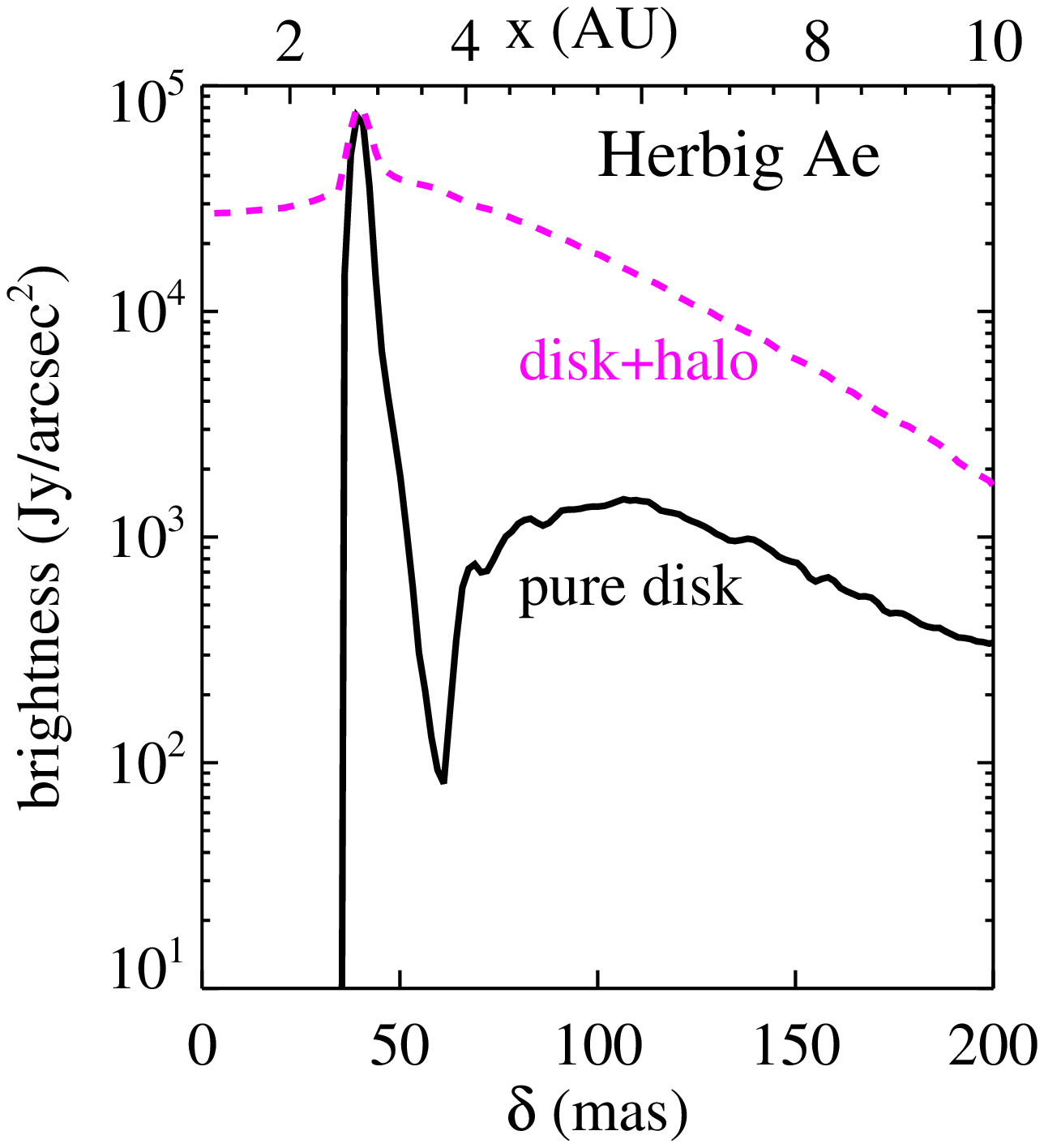}
\end{center}
\caption{{\it {Top}}: Mid--IR image of a disk heated by a Herbig Ae
  star at distance of 50\,pc and viewed at $30^{\circ}$. Color bar is
  in $\log($Jy/arcsec$^2)$. {\it {Bottom}}: Mid--IR surface brightness
  distribution measured as a cut along $\delta$ through the origin of an
  emission image of a pure disk (full line) and that computed in a
  disk plus halo configuration (dashed).  \label{HEBEima.ps}}
\end{figure}

Disks of the more luminous Herbig Ae stars have an evaporation zone
farther out than around T Tauri stars. At these distances the
gravitation is reduced and the disks become thicker than for T Tauri
disks. The height of the Herbig Ae disks and a shadowed presentation
is displayed in Fig.~\ref{HEBEsurf.ps}. When compared to T Tauri stars
(Fig.~\ref{TTSsurf.ps}), the ratio $z_{\rm{bot}}(r)/r$ is indeed
higher and has less structure. There is a pronounced puffed--up inner
rim causing a far reaching shadow, so that a second one farther away
than 30AU is only envisaged. However, this far away the midplane
temperature has dropped to below 30\,K (Fig.~\ref{halo.ps}) and other
effects than stellar heating may dominate the disk structure. We
display the mid--IR image and surface profile of the disk heated by a
Herbig Ae star in Fig.~\ref{HEBEima.ps}. At least one striking gap and
a ring in the disk are identified.


\section{Conclusion}

The infrared appearance of passive disks around low and medium mass
stars has been discussed. Dust particles are made either of bare
material or fluffy composites of silicate and carbon. Grain sizes
range from 160\AA \/ to 3000\AA \/ as derived for the ISM and up to
33$\mu$m assuming grain growth in the disks. The disks are heated by
stellar light with luminosities between 2\,\Lsun \/ and 50\,\Lsun
\/. The disks have an inner radius that is roughly determined by the
dust evaporation temperature of the grain material. The surface
density of the disks was set close to what is estimated for the
minimum mass of the solar nebula. Disks are configured with or without
halos. Two hydrostatic and geometrically thin disk scenarios were
examined for which we assume that gas and dust are mixed and have the
same temperature.

i) In {\it {1+1D}} the disks were divided in small ring segments in
which the radiation transfer was solved.  For each ring, a slab
geometry was applied, and it was assumed that in deeper layers the
disk is isothermal.  The transport of radiation in the radial
direction was ignored.  From the surface of 1+1D disks, the star is
always visible.  A flat or flaring structure of the disk was
considered. This type of disk is in widespread use in the literature.

ii) In {\it {2D disks}} the hydrostatic and radiative transfer
equations were solved in an iterative scheme.  For the latter an
accurate MC method was presented that can be applied to arbitrary dust
geometries.  We used an adaptive three--dimensional Cartesian grid
where cubes are divided into subcubes whenever required, for example,
close to the disk surface. The MC method is vectorized and the
parallelization is realized to work on graphics cards or conventional
processing units and provides a speed--up roughly proportional to the
number of processor cores or parallel threads, available.  On our
conventional computer, the vectorized code is a factor 100 faster than
the scalar version. In regions of very high optical depth, for example
close to the midplane of the disk, photon packets may get trapped. In
this case the algorithm is further accelerated by a modified random
walk procedure. The 2D disks provide a more realistic description than
the 1+1D disks: In the RT solution of the 1+1D disks it is assumed
that the radial transport of radiation can be ignored and that layers
with vertical optical of $\tau_{\rm {top}} \simgreat 20$ are
isothermal. These assumptions break for example near the inner regions
of the disk. On the other hand, in the 2D disks the radial transport
of radiation is included and no assumptions are made that regions of
the disks need to be isothermal.  Therefore a self consistent
treatment of the hydrostatic balance is only provided in the 2D
disks. Our main findings are that:

\begin{itemize}

\item 1+1D disks have a smooth surface structure, and their midplane
  temperature is approximated by a power law well. They gravely
  underestimate the mid IR emission when compared to 2D disks.

\item Halos substantially increase the IR emission and the temperature
  of the midplane, which is smoother than obtained in disks without
  halos.

\item In 2D disks the midplane temperature shows up and downs with
  radius (Fig.\ref{sedIt.ps}). Such a hilly structure is also
revealed in their surface structure. 

\item Emission images of 2D disks display gaps and ring--like
  structures in particular in the region of terrestrial planets.  Disk
  structures are caused by puffing up the disk surface and followed up
  by their shadows behind. The disk structure is dimmed when halos are
  considered. Rings and gaps are more pronounced in T Tauri stars than
  in Herbig Ae stars.

\end{itemize}

The detected gap and ring structures of the disks are only caused by
the assumption of hydrostatic equilibrium and the derived detailed
vertical temperature profiles of the disk.  On the other hand, it is
easy to imagine, without the need of such detailed computations, that
behind a puffed--up disk the heating is dimmed and therefore that the
dust temperatures are lower. This causes that in those shadowed
regions the height of the disk is smaller unless it is again exposed
to direct stellar light.  Therefore the detected gap and ring--like
structures are a plausible effect of the shadows, but they have not
yet been reported by others. We feel that this is important because
ring--like structures of protoplanetary disks are often interpreted as
``fingerprints'' of the planet formation process, whereas in the
models presented planets have not been considered.

\begin{acknowledgements} 
{We are grateful to Endrik~Kr\"ugel for discussions and helpful
  suggestions.}
\end{acknowledgements}

\end{document}